\documentclass[aps,prb,preprint,showkeys,showpacs,floatfix]{revtex4}

\usepackage{graphicx,color}
\usepackage{natbib}

\usepackage{float}
\graphicspath{{figs/}}
\begin{document}

\title{Modulation of thermal conductivity in single-walled carbon nanotubes by fullerene encapsulation: enhancement or reduction?}

\author{Jing Wan}
    \affiliation{Shanghai Institute of Applied Mathematics and Mechanics, Shanghai Key Laboratory of Mechanics in Energy Engineering, Shanghai University, Shanghai 200072, People's Republic of China}

\author{Jin-Wu Jiang}
    \altaffiliation{Corresponding author: jwjiang5918@hotmail.com}
    \affiliation{Shanghai Institute of Applied Mathematics and Mechanics, Shanghai Key Laboratory of Mechanics in Energy Engineering, Shanghai University, Shanghai 200072, People's Republic of China}

\date{\today}
\begin{abstract}

Fullerence encapsulation has been proven to be a powerful approach to enhance mechanical and electronic properties of the single-walled carbon nanotubes (SWCNTs). However, some discrepancies emerge in recent studies of the fullerence encapsulation effect on the thermal conductivity of SWCNTs. More specifically, most previous theoretical works predicted slightly enhancement of the thermal conductivity, but the recent experiment by Kodama {\it et al.} (Nat. Mat. \textbf{16}, 892 (2017)) observed clear reduction of the thermal conductivity in SWCNTs by fullerence encapsulation. We perform molecular dynamics simulations to revisit this issue, by comparatively investigate the thermal conductivity of the SWCNT (n, n) and the corresponding peapod (n, n) with n = 8, 9, 10, and 11. We find that the fullerence encapsulation can reduce the thermal conductivity of narrower SWCNTs with n = 8 and 9, but it can slightly enhance the thermal conductivity of thicker SWCNTs with n = 10 and 11. The underlying mechanisms for these opposite effects are explored by analyzing the encapsulation induced structural deformation and the variation of the phonon dispersion of the SWCNT. We illustrate that the reduction effect observed in the recent experiment is related to the mechanism for reduction of the thermal conductivity for narrower SWCNTs here, while previous numerical works correspond to the enhancement effect for thicker SWCNTs found in the present work. Our findings shed lights on clarifying the discrepancy between previous numerical predictions and experimental observations on the effect of the fullerence encapsulation on the thermal conductivity of the SWCNT.

\end{abstract}

\keywords{Carbon naotube, Fullerene encapsulation, Thermal conductivity}
\pacs{61.46.+W, 61.48.+c, 65.80.+n}
\maketitle
\pagebreak

\section{Introduction}

Single-walled carbon nanotubes (SWCNTs) have attracted significant research interests for their remarkable properties.\cite{Baughman2002,Endo2007} From the mechanical point of view, SWCNTs' characteristic hollow structure may be a disadvantage for their practical applications, as such structure has weak ability to resist the deformation in the radial direction, especially for SWCNTs of large diameters.\cite{ChangT2008prl} The encapsulation of some molecules would be a direct way to upgrade SWCNT's mechanical property.\cite{Khlobystov2005} One such example is the encapsulation of the fullerence C$_{60}$ into the SWCNT, forming the so-called peapod structure, which is the subject discussed in the present work. Indeed, simulations have shown that the encapsulation of the fullerence C$_{60}$ can greatly improve the mechanical strength of the SWCNT to resist the radial deformation.\cite{Wu2011} Besides the enhancement of the mechanical properties, previous works have disclosed the effect of the fullerence encapsulation on the electronic properties for the SWCNT.\cite{hornbaker2002,Pfeiffer2004,Zhu2007,Okazaki2008,Wu2011} Okada et al. found that the peapod also has some novel electronic properties,\cite{Okada2001} while the scanning tunneling microscope measurement observed the modulation of the local electronic structure of the nanotube by the fullerence encapsulation.\cite{Okazaki2008} 

Some previous works have been devoted to examining the effect of the fullerence encapsulation on the thermal conductivity of the SWCNT. Molecular dynamics (MD) simulations found that the fullerence encapsulation can enhance the thermal conductivity of the SWCNT due to low-frequency radial vibration coupling between fullerenes and nanotube and fullerene-fullerene collisions.\cite{Noya2004} The enhancement of thermal conductivity by the fullerence encapsulation has also been predicted in other theoretical works.\cite{Kawamura2008,Cui2015} However, quite different results are observed in the experiment. In 2002, Vavro et al. carried out experiments to demonstrate that the fullerence encapsulation has almost no effect on the thermal conductivity of the SWCNT, because the encapsulation results in two competing effects, i.e. the increase of phonon-phonon scattering and the introduction of new phonon conduction channels.\cite{Vavro2002} More recently, Kodama et al. observed clear reduction of the thermal conductivity for the SWCNT by the encapsulation of the fullerence C$_{60}$.\cite{Kodama2017} An important task in this field is thus to resolve the discrepancy between the existing theoretical predictions and the available experiments on the fullerence encapsulation effect on the thermal conductivity of the SWCNT.

In this paper, we perform MD simulations to examine the effect of the fullerence encapsulation on the thermal conductivity of the SWCNT. We find that the fullerence encapsulation can greatly reduce the thermal conductivity of the narrower SWCNT (n, n) with n = 8 and 9, which is attributed to the increased phonon-phonon scattering and the decreased acoustic velocities due to the strong interaction between the outer nanotube shell and the encapsulated fullerence. In contrast, the fullerence encapsulation slightly enhance the thermal conductivity of the thicker SWCNT (n, n) with n = 10 and 11, owing to the additional phonon conduction channels contributed by the encapsulated fullerences. Our simulations may assist to the clarify of the discrepancy between previous numerical predictions and experimental observations.

\section{Structure and Simulation Details}

\begin{figure}[tb]             
  \begin{center}
    \scalebox{1.0}[1.0]{\includegraphics[width=8cm]{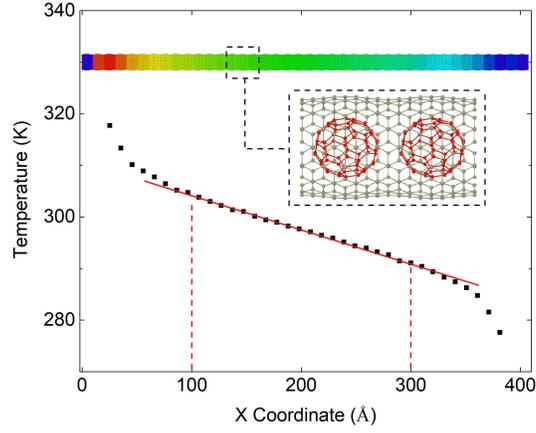}}
  \end{center}
  \caption{Temperature profile for the MD simulation of the thermal transport in the peapod (8, 8). Peapod (n, n) is constructed by the encapsulation of fullerences C$_{60}$ into the SWCNT (n, n). The temperature profile in the middle region is linearly fitted to extract the temperature gradient $\frac{dT}{dx}$. Inset shows the spatial distribution of the temperature in the peapod.}
  \label{fig_dTdx}
\end{figure}

We consider four SWCNTs (n, n) with n = 8, 9, 10, and 11 in the present work. The peapod (n, n) is constructed by the encapsulation of fullerences C$_{60}$ into the SWCNT (n, n), as depicted by the inset of Fig.~\ref{fig_dTdx}. These peapods simulated here contain forty fullerences. The peapod is divided into forty sections in the simulation of the thermal transport, while the SWCNT is also divided into fourty sections accordingly. The 1st and last sections are fixed during the MD simulation, and the thermal energy is pumped into (out) the system in the 3rd (38th) section using the Nos\'e-Hoover thermostat.\cite{Nose,Hoover}.

All MD simulations were performed using the large-scale atomic molecular massively parallel simulator (LAMMPS) package,\cite{PlimptonSJ} while the OVITO package is used for visualization.\cite{ovito} The covalent bonding among carbon atoms is described by the optimized Tersoff potential.\cite{Lindsay2010} The van der Waals interaction is described by the Lennard-Jones potential with parameters $\varepsilon = 2.4$~{meV} and $\sigma = 3.4$~{\AA}.\cite{Kawamura2008} The standard Newton equations of motion were integrated in time using the velocity Verlet algorithm with a time step of 0.5~fs.

\begin{figure}[tb]             
  \begin{center}
    \scalebox{1.0}[1.0]{\includegraphics[width=8cm]{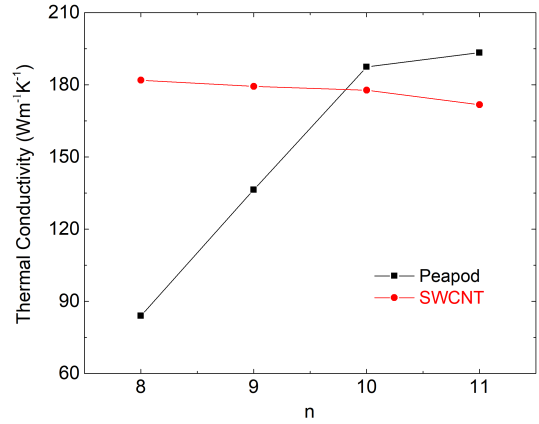}}
  \end{center}
  \caption{A comparison between the thermal conductivity for the SWCNT (n, n) and peapod (n, n) at room temperature.}
  \label{fig_conductivity}
\end{figure}

\section{Results and Discussions}

\subsection{Simulation results}

The thermal transport was simulated using the non-equilibrium MD approach. The thermal conductivity $\kappa$ is obtained from the Fourier law,
\begin{eqnarray}
J =-\kappa (\frac{dT}{dx}).
\label{eq_fourier}
\end{eqnarray}
where $J$ is the non-equilibrium steady-state heat flux density, and $\frac{dT}{dx}$ is the temperature gradient along the tube axis direction. Fig.~\ref{fig_dTdx} shows one sampling temperature profile at room temperature, which is for the peapod (8, 8). The temperature profile in the middle region is linearly fitted to extract the value of $\frac{dT}{dx}$.

Figure~\ref{fig_conductivity} compares the thermal conductivity of the SWCNT (n, n) and peapod (n, n) at room temperature. For n = 8 and 9, the thermal conductivity of the peapod is considerably smaller than the SWCNT; i.e., the fullerence encapsulation can effectively reduce the thermal conductivity of the SWCNT (n, n) for n = 8 and 9. This result agrees quite well with the recent experimental measurement, which observed that the fullerence encapsulation strongly reduces the thermal conductivity of the SWCNT.\cite{Kodama2017}

In contrast, for n = 10 and 11, the thermal conductivity of the peapod is larger than the SWCNT, but the difference is small. In other words, the fullerence encapsulation is able to slightly enhance the thermal conductivity of the SWCNT (n, n) for n = 10 and 11. This result is consistent with previous theoretical calculations. For example, Noya et al. found that the fullerence encapsulation can enhance the thermal conductivity of the SWCNT due to low-frequency radial vibration coupling between fullerenes and nanotube and fullerene-fullerene collisions.\cite{Noya2004} The enhancement of thermal conductivity by the fullerence encapsulation has also been predicted in other works.\cite{Kawamura2008,Cui2015}

Our simulations thus explore two regions with opposite modulation effects on the thermal conductivity by the fullerence encapsulation. We will illustrate in the rest of this paper that the recent experimental observations are closely related to the region with n = 8 and 9 in Fig.~\ref{fig_conductivity}, while the previous theoretical works correspond to the region with n = 10 and 11.

\begin{figure*}[tb]             
  \begin{center}
    \scalebox{1.0}[1.0]{\includegraphics[width=\textwidth]{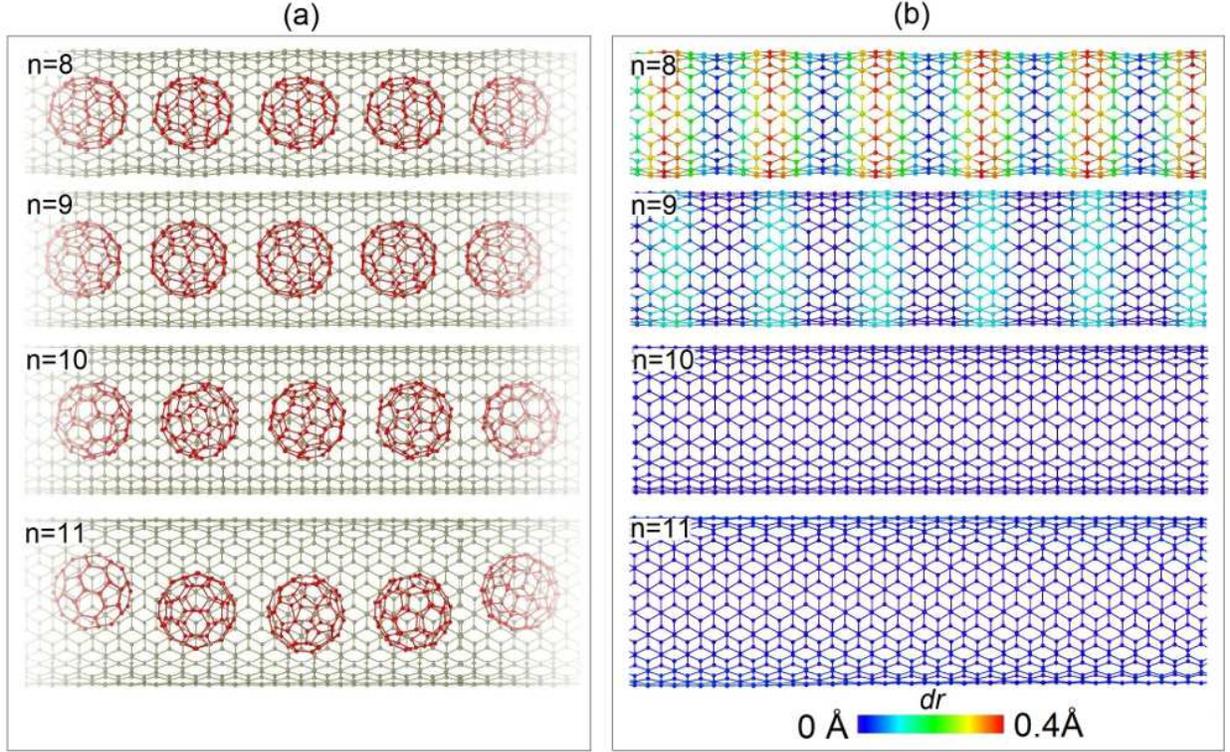}}
  \end{center}
  \caption{Encapsulation induced deformation in the peapod. (a) Optimized configuration of the peapod with n = 8, 9, 10, and 11 from top to bottom. (b) Variation for the radius of the outer nanotube shell in the peapod (n, n) with n = 8, 9, 10, and 11 from top to bottom. Note obvious encapsulation induced periodic deformation of the outer narrower nanotube shells for n = 8 and 9. Color represents the variation of the radius ($dr$) with respective to the radius of the corresponding SWCNT.}
  \label{fig_peapod_dr}
\end{figure*}

\begin{figure}[tb]             
  \begin{center}
    \scalebox{1.0}[1.0]{\includegraphics[width=8cm]{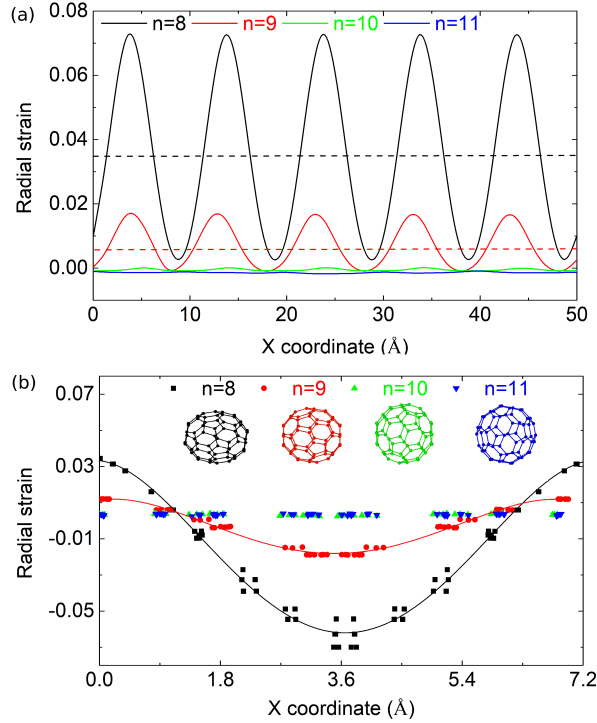}}
  \end{center}
  \caption{The radial strain distribution in the peapod (n, n). (a) The radial strain distribution for the outer nanotube shell in the peapod. (b) The radial strain distribution for the encapsulated fullerence in the peapod. Note obvious deformation of the encapsulated fullerences from sphere into ellipsod in the peapod (8, 8) and peapod (9, 9). The x-direction is along the axial direction of the nanotube.}
  \label{fig_strain}
\end{figure}

\subsection{Structural analysis}

We first present some structural analysis for the effect of the fullerence encapsulation on the thermal conductivity of the SWCNT. Fig.~\ref{fig_peapod_dr} shows the optimized configuration of one segment in the peapod (n, n) with n = 8, 9, 10, and 11. Fig.~\ref{fig_peapod_dr}~(a) illustrates that the fullerences are tightly encapsulated inside the narrower SWCNT with n = 8 and 9. As a result of the strong encapsulation, there are some periodic expansion in the outer nanotube shell and the fullerences are arrayed straightly inside the SWCNT. The periodic expansions are most obvious in the peapod (8, 8). The resultant radial strains for the outer nanotube shell and the encapsulated fullerence are shown in Fig.~\ref{fig_strain}, which demonstrates obvious deformations in the radial direction of the outer nanotube shell and the encapsulated fullerence in the peapod (8, 8) and (9, 9). The strong interaction between the outer nanotube shell and the encapsulated fullerence can cause phonon-phonon scattering for the thermal transport, leading to the reduction of the thermal conductivity of the SWCNT by fullerence encapsulation.\cite{Kodama2017} Furthermore, the periodic radial strain distribution results in additional interface-like scattering mechanism for the thermal transport, which results in further reduction of the thermal conductivity.\cite{Kodama2017} As a result, the fullerence encapsulation can reduce the thermal conductivity of the SWCNT (8, 8) and (9, 9).

For the peapod (n, n) with n = 10 and 11, the fullerence encapsulation does not cause any radial deformations in the SWCNT as shown in Figs.~\ref{fig_peapod_dr} and \ref{fig_strain}. It is because the space between the outer nanotube shell and the encapsulated fullerence is large, so the van der Waals interaction between them is weak. As a result, the fullerence encapsulation does not bring strong phonon-phonon scatterings and will not lead to obvious effects on the thermal conductivity of the SWCNT for n = 10 and 11.

\subsection{Lattice dynamics analysis}

\begin{figure}[tb]             
  \begin{center}
    \scalebox{1.0}[1.0]{\includegraphics[width=8cm]{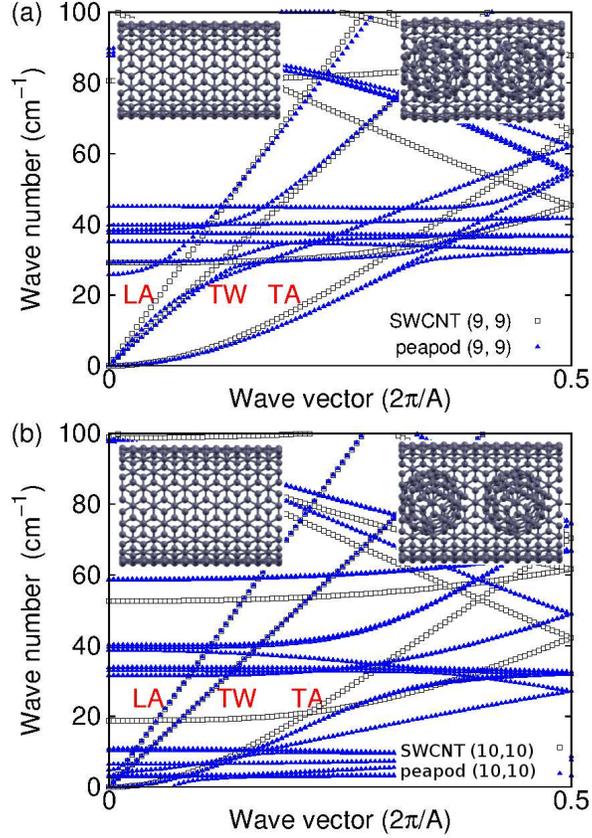}}
  \end{center}
  \caption{Comparison between the phonon dispersions of the SWCNT and peapod. (a) Low-frequency phonon dispersions for SWCNT (9, 9) and peapod (9, 9). (b) Low-frequency phonon dispersions for SWCNT (10, 10) and peapod (10, 10). Insets show the unit cell used for the phonon calculation of the SWCNT and peapod. The length of the unit cell is $A=8a$, with $a=2.49$~{\AA} as the lattice constant of graphene. The four acoustic branches are denoted by LA, TW, and TA (doubly degenerate).}
  \label{fig_phonon}
\end{figure}

The lattice thermal conductivity studied in the above MD simulations is contributed by the phonon modes of the SWCNT or the peapod. The phonon dispersion can thus provide some fundamental information for the effect of the fullerence encapsulation on the thermal transport of the SWCNT. In particular, acoustic phonons play an important role in the lattice thermal transport phenomenon, and the overall thermal conductivity is proportional to the group velocities of acoustic phonons, according to the kinetic theory. Hence, we will focus on the fullerence encapsulation effect on the low-frequency phonons in the SWCNT. To this end, we compare the phonon dispersion in the low-frequency range [0, 100]~{cm$^{-1}$} for the SWCNT and the peapod in Fig.~\ref{fig_phonon}.

Figure~\ref{fig_phonon}~(a) shows the phonon dispersion for the SWCNT (9, 9) and the peapod (9, 9). For convenience of the comparison between SWCNT and peapod, considerably large unit cells (top left and rigth insets) have been used in the phonon calculation. The length of the unit cell along the axial direction is $A=8a$, with $a=2.49$~{\AA} as the lattice constant for graphene.\cite{Lindsay2010} There are 288 carbon atoms in the unit cell of the SWCNT (9, 9), while there are in total $288+2\times 60=408$ carbon atoms in the peapod (9, 9). There are four acoustic phonon branches in the tubal SWCNT and peapod, i.e., the longitudinal acoustic (LA) branch, the twisting (TW) branch, two doubly degenerate transverse acoustic (TA) branches. As a result of the fullerence encapsulation, there are some additional phonon branches around 40~{cm$^{-1}$} in the peapod (9, 9), which correspond to the vibration of the fullerence.

\begin{table}
\caption{Group velocities for the LA and TW acoustic branches in the SWCNT and the peapod. The coefficient $\alpha$ is for the flexural TA branch, which has a dispersion of $\omega=\alpha k^2$ with $k$ as the wave vector. The ratio of the group velocities for the peapod and the SWCNT is shown in the fourth line and the last line.}
\label{tab_vg}
\begin{tabular}{|c|c|c|c|}
\hline 
 & $v_{\rm LA}$ (kms$^{-1}$) & $v_{\rm TW}$ (kms$^{-1}$) & $\alpha$ ($\AA$kms$^{-1}$)\tabularnewline
\hline 
\hline 
SWCNT (9, 9) & 21.6 & 15.7 & 80.3\tabularnewline
\hline 
peapod (9, 9) & 18.0 & 13.2 & 65.8\tabularnewline
\hline 
peapod/SWCNT & 0.83 & 0.84 & 0.82\tabularnewline
\hline 
\multicolumn{1}{|c}{} & \multicolumn{1}{c}{} & \multicolumn{1}{c}{} & \tabularnewline
\hline 
SWCNT (10, 10) & 21.9 & 14.9 & 97.1\tabularnewline
\hline 
peapod (10, 10) & 21.9 & 14.9 & 88.0\tabularnewline
\hline 
peapod/SWCNT & 1.0 & 1.0 & 0.91\tabularnewline
\hline 
\end{tabular}
\end{table}

A distinct effect of the fullerence encapsulation is the reduction of the group velocities of these four acoustic phonon branches; i.e., the group velocities for the four acoustic branches of the peapod (9, 9) are considerably smaller than that of the SWCNT (9, 9). The group velocities for the LA and TW branches are shown in Tab.~\ref{tab_vg}. It should be noted that the TA phonon branch is a flexural mode, owing to the rod-like structure of the SWCNT or the peapod.\cite{LandauLD} As a result, the phonon frequency $\omega$ depends on the wave vector $k$ as a parabolic function, i.e., $\omega=\alpha k^2$. The coefficient $\alpha$ is shown in the last row of Tab.~\ref{tab_vg}. The fourth line in Tab.~\ref{tab_vg} shows the ratio between the group velocity of the peapod and the SWCNT. This ratio is smaller than 1.0, indicating the reduction of the group velocities and the coefficient $\alpha$.

The reduction of the group velocities and the coefficient $\alpha$ can be explained by the variation of the mass density caused by the fullerence encapsulation. For narrower SWCNTs like (9, 9), the van der Waals interaction between the outer nanotube shell and the fullerence is strong. For instance, the Lennard-Jones potential for the peapod (9, 9) is 10.58~{eV}. This positive potential value indicates that the relaxed space between the outer nanotube shell and the encapsulated fullerence is under the compressive condition. The periodic expansion of the outer nanotube shell shown in the above Fig.~\ref{fig_peapod_dr} also implies the strong interaction between the outer nanotube shell and the fullerence in the peapod (9, 9). The peapod can thus be regarded as an integrated rigid body, since the outer nanotube shell and the encapsulated fullerence is strongly interacted. In this sense, the effect of the fullerence encapsulation is to increase the mass density of the SWCNT, assuming that the encapsulation induced variation in the volumn of the SWCNT is neglectable. The ratio between the mass density of the peapod ($\rho_{p}$) and SWCNT ($\rho_{s}$) is $\rho_{\rm s}/\rho_{\rm p}=288/408=0.71$. The group velocities ($v$) for the acoustic phonons and the coefficient $\alpha$ for the flexural phonon are inversely proportional to the square of the mass density.\cite{LandauLD} As a result, the ratio between the velocity or the coefficient of the peapod ($v_p$) and SWCNT ($v_s$) is $v_p/v_s=\sqrt{\rho_{\rm s}/\rho_{\rm p}}=0.84$ or $\alpha_p/\alpha_s=0.84$. The ratios of the group velocity and the coefficient $\alpha$ in Tab.~\ref{tab_vg} are all very close to 0.84 for the peapod (9, 9) and the SWCNT (9, 9). This excellent agreement informs that the fullerence encapsulation leads to the enhancement of the mass density of the SWCNT (9, 9), which results in the reduction of the group velocities of all acoustic phonons. As a result, the peapod (9, 9) has a smaller thermal conductivity than the SWCNT (9, 9). In other words, the thermal conductivity is considerably reduced by fullerence encapsulation for narrower SWCNTs, because the acoustic velocities are considerably reduced. Similar mechanism is applicable to explain the reduction of the thermal conductivity of the SWCNT (8, 8) by fullerence encapsulation, as the interaction between the outer nanotube shell and the encapsulated fullerence is also strong in the peapod (8, 8).

\begin{figure}[tb]             
  \begin{center}
    \scalebox{1.0}[1.0]{\includegraphics[width=6cm]{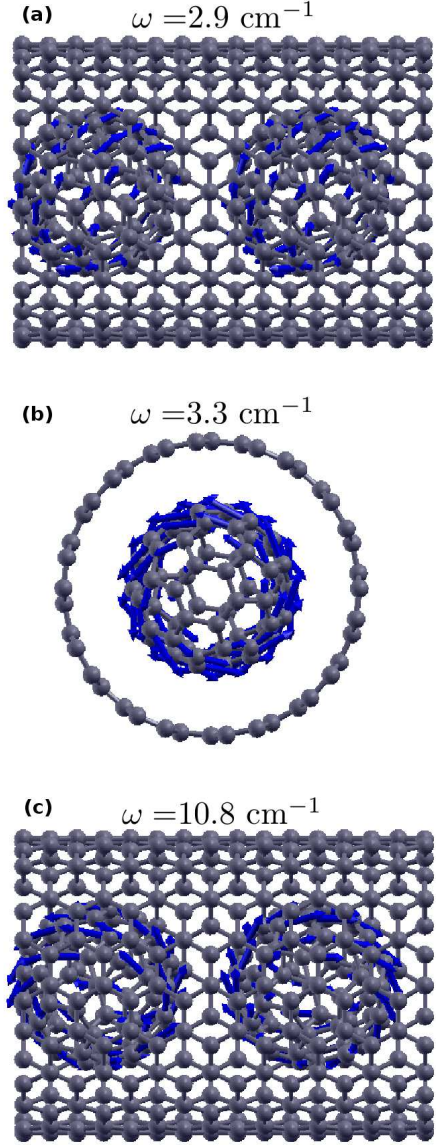}}
  \end{center}
  \caption{Vibrational morphology for the quasi-acoustic phonon modes of the peapod (10, 10) at the $\Gamma$ point $k=0$. The frequencies are (a) 2.9~{cm$^{-1}$}, (b) 3.3~{cm$^{-1}$}, and (c) 10.8~{cm$^{-1}$}. Only fullerenes are involved in the vibration of these quasi-acoustic modes. The arrow attached to the atom represents the vibrational component of each atom in the phonon mode.}
  \label{fig_u}
\end{figure}

Figure~\ref{fig_phonon}~(b) compares the low-frequency phonon dispersions of the SWCNT (10, 10) and peapod (10, 10). Insets show the unit cells used for the phonon calculation. In contrast to the (9, 9) results shown in Fig.~\ref{fig_phonon}~(a), the group velocities for the LA and TW branches are almost the same in the peapod (10, 10) and SWCNT (10, 10); i.e., these two group velocities are not affected by the fullerence encapsulation for the SWCNT (10, 10). The coefficient $\alpha$ for the TA branch is only slightly reduced by the fullerence encapsulation. The fullerence encapsulation effect on the acoustic velocities of the SWCNT (10, 10) are also tabulated in Tab.~\ref{tab_vg}. The encapsulation insensitivity of the acoustic velocities in the SWCNT (10, 10) is due to the weak van der Waals interaction between the outer (10, 10) nanotube shell and the encapsulated fullerence. The Lennard-Jones potential between the outer nanotube shell and the fullerence is -6.05~{eV}. The negative Lennard-Jones potential energy implies that there is no compression for the space between the outer nanotube shell and the encapsulated fullerences. The van der Waals interaction is small as compared with the the covalent bonding within the SWCNT, so the fullerence encapsulation has neglectable effects on the velocities of the LA and TW branches whose vibrations are governed by the covalent bonding. The fullerence encapsulation causes some weak effects on the coefficient $\alpha$ of the flexural TA branches, because these flexural modes are related to the bending energy of the rod-like SWCNT that is also on the small energy level.

The second distinct feature of Fig.~\ref{fig_phonon}~(b) is the emergence of a series of quasi-acoustic phonon branches in the peapod (10, 10), which have frequencies lower than 10~{cm$^{-1}$}. The phrase `quasi-acoustic' means that these branches have ultra-low frequencies and small finite acoustic velocities. These phonons correspond to the vibration of the encapsulated fullerence, while the outer nanotube shell is not involved, due to the weak interaction between the outer nanotube shell and the encapsulated fullerence. Fig.~\ref{fig_u} displays the vibrational morphology of three representative phonons at the $\Gamma$ point with $k=0$. The vibration of these phonons only involves the encapsulated fullerences, which are interacted through the weak van der Waals interaction, so these phonons have low frequencies and small group velocities. In other words, the encapsulated fullerences provide additional channels to transport the thermal energy, which tends to enhance the thermal conductivity of the SWCNT (10, 10).

We thus find that for thicker SWCNTs like (10, 10), the fullerence encapsulation induces only weak effects on the acoustic velocities, but it brings additional thermal transport channels. As a consequence, the thermal conductivity is enhanced by the fullerence encapsulation for SWCNT (10, 10). These analyses here are also applicable for the fullerence encapsulation effect on the thermal conductivity of SWCNT (11, 11).

\section{Conclusion remarks}

As a conclusion remark, we comment on the relation between our results and previously published theoretical and experimental results. There are two major results in the present work.

First, the thermal conductivity of the SWCNT (n, n) is greatly reduced by fullerence encapsulation for n = 8 and 9. It is because the strong interaction between the outer nanotube shell and the encapsulated fullerences increases the phonon-phonon scattering and decreases the acoustic phonon velocities. In the recent experiment,\cite{Kodama2017} the diameter of the peapod is $1.350 \pm 0.057$, which is probably corresponding to the SWCNT (10, 10), not SWCNT (9, 9). According to our simulations and previous simulations, the fullerence encapsulation induced deformation shall be very small for the SWCNT (10, 10).\cite{Noya2004,Kawamura2008,Cui2015} However, in the experiment, they observed clearly considerable radial strains in the peapod samples. It may indicate that, besides the fullerence C$_{60}$, some other atoms or functional groups are also encapsulated inside the SWCNT, which results in the deformation of the SWCNT (10, 10). Nevertheless, the obvious radial deformation in the experimental peapod samples implies a strong interaction between the outer nanotube shell and the encapsulated fullerence in the peapod, which results in the reduction of the thermal conductivity for the SWCNT by the fullerence encapsulation, according to the discussions of our first results.

Second, the thermal conductivity of the SWCNT (n, n) is only slightly enhanced by fullerence encapsulation for n = 10 and 11. It is because the weak interaction between the outer nanotube shell and the encapsulated fullerence does not cause obvious phonon-phonon scattering or the reduction of the acoustic velocities. On the contrary, the encapsulated fullerences provide some additional weak channels (with small group velocities) for the lattice thermal transport. As a result, the thermal conductivity is slightly enhanced by fullerence encapsulation for the SWCNT (10, 10) and (11, 11). This result is in consistence with previous numerical simulations.\cite{Noya2004,Kawamura2008,Cui2015}

\textbf{Acknowledgements} The work is supported by the Recruitment Program of Global Youth Experts of China, the National Natural Science Foundation of China (NSFC) under Grant No. 11504225, and the Innovation Program of Shanghai Municipal Education Commission under Grant No. 2017-01-07-00-09-E00019.

\bibliographystyle{aipnum4-1} 
\bibliography{biball}

\end{document}